\documentclass[aps,prd,preprint,showpacs,nofootinbib]{revtex4}
\usepackage{latexsym,bm,amsmath,amssymb}
\DeclareMathAlphabet{\mathpzc}{OT1}{pzc}{m}{it}
\def\lng{\mbox{ln$|g|$}}
\begin{document}

\title{Covariant Symplectic Structure and Conserved Charges of New Massive Gravity}
\author{G{\"o}khan Alka\c{c}}
\email{alkac@metu.edu.tr}
\author{Deniz Olgu Devecio\u{g}lu}
\email{dedeveci@metu.edu.tr}
\affiliation{Department of Physics, Faculty of Arts and  Sciences,\\
             Middle East Technical University, 06800, Ankara, Turkey}

\date{\today}

\begin{abstract}
We show that the symplectic current obtained from the boundary term,
which arises in the first variation of a local diffeomorphism
invariant action, is covariantly conserved for any gravity theory
described by that action. Therefore, a Poincar\'{e} invariant two-form
can be constructed on the phase space, which is shown to be closed
without reference to a specific theory. Finally, we show that one
can obtain a charge expression for gravity theories in various
dimensions, which plays the role of the Abbott-Deser-Tekin 
charge for spacetimes with nonconstant curvature backgrounds, by
using the diffeomorphism invariance of the symplectic two-form. As an
example, we calculate the conserved charges of some solutions of new
massive gravity and compare the results with previous
works.

\end{abstract}

 \maketitle
\section{Introduction}
The inherent nature of Hamiltonian formulation seems to shelter a
conflict with the sacred property of general covariance by the
choice of time coordinate. This very fact is a major obstacle in the
definition of conserved charges especially in gravity theories. The
Arnowitt-Deser-Misner (ADM) mass \cite{adm} is a good example of this hindrance. One has to
deal with the decomposition of spacetime into spacelike
hypersurfaces parametrized by the time coordinate. Achieving this
for higher curvature gravity theories is obviously a tedious task to
perform.

The aim of this paper is to circumvent this difficulty by employing
the construction of \cite{got,wit,zuck} which simply builds up the
phase space from the solutions of the classical equations. The
symplectic two-form identified through this way contains all of the
relevant properties of the phase space without the need for defining
momenta. Having constructed the symplectic structure, the
diffeomorphism invariance of the symplectic two-form lets one find a
closed expression to compute the conserved charges of the solutions
of the theory, which is of paramount importance to understand the
thermodynamical properties of the solutions. The most important
result we will prove in this paper is the equivalence of this charge
expression to the Abbott-Deser-Tekin (ADT) \cite{des1,des2,des3} charge when the
diffeomorphisms are restricted to be the isometries of the
background spacetime.

For topologically massive gravity (TMG) \cite{djt} the symplectic
two-form and the conserved charges were given in \cite{bayramabi}. In
this work we focus on a three dimensional gravitational theory that
has attracted considerable attention recently. This theory, termed
as new massive gravity (NMG) \cite{hohm,hohm2}, is obtained by
adding a particular higher curvature term ($\alpha R^{2}+ \beta
R_{ab}^{2}$ with the constraint $8\alpha+3 \beta=0$) to the
Einstein-Hilbert action, which makes the theory tree-level unitary
\cite{ibo} but not renormalizable \cite{mun}. It is a valuable toy
model for our purposes since many interesting solutions with
AdS$_{3}$ and arbitrary backgrounds have appeared in the literature
\cite{btz,clem3,ott,giri1,clem}.

The outline of the paper is as follows: Sec. \ref{construct}
starts with the  definition of the fundamental objects on the phase
space and continues with the construction of the symplectic two-form
$\omega$, for the theories derived from the action
\[ I = \int \, d^{D}x \, \sqrt{|g|} \, \Big( \frac{1}{\kappa} (R + 2 \Lambda_{0})
+ \alpha R^{2} + \beta R_{ab}^{2}  \Big).\] We end up the section
with the discussion of the gauge invariance of $\omega$. In Sec.
\ref{conserved}, we find an expression for the conserved charges of
these theories and show its equivalence to the ADT charge. Section
\ref{soln} is devoted to the computation of the energy and angular
momentum of some solutions of NMG using the formulas derived in
Sec. \ref{conserved}.

Our conventions are as follows: The signature of the metric is
$(-,+,\cdots,+)$. The Riemann tensor is defined through
$[\nabla_{a},\nabla_{b}]V_{c}=R_{abc}\,^{d}V_{d}$ and $ R_{ab}:=
R^{c}\,_{acb}$. For the symmetrization and antisymmetrization of
tensors, the factors and signs are chosen so that e.g.
$T_{(ab)}\equiv \frac{1}{2}(T_{ab}+T_{ba})$, $T_{[ab]}\equiv
\frac{1}{2}(T_{ab}-T_{ba})$.

\section{\label{construct}The construction of the symplectic two-form}
First we start by summarizing the covariant canonical formulation of
classical theories developed by \cite{got,wit,zuck}, in a way that
manifestly preserves relevant symmetries of the theory. Before
delving into details, let us recall the usual canonical formalism of
a theory. One starts with a $2N$ dimensional smooth manifold $Z$
endowed with a two-form given as
\begin{equation}
\omega=dp_{i}\wedge dq^{i},
\end{equation}
where $q^{i}$ and $p_{i}$ are introduced as generalized coordinates
and generalized momenta, respectively, and $i=1,\dots N$.
Furthermore, $\omega$ is closed ($d \omega= 0$) and nondegenerate,
i.e. when $\omega$ is written as a $2N \times 2N$ matrix, it has an
inverse. This closed two-form $\omega$ on $Z$ is called the symplectic
two-form.

In order to develop and use this structure in geometrical theories
derived from an action, we need to follow a somewhat different route
from the usual approach discussed above, since choosing $p_{i}$ and
$q^{i}$ as coordinates of the phase space $Z$ would destroy the
general covariance (by the choice of time coordinate). One should
construct the phase space $Z$ from the solutions of the equations of
motion to achieve a manifestly covariant structure. Since classical
solutions of any physical theory are in one-to-one correspondence
with the initial values of $p_{i}$ and $q^{i}$, we define our phase
space as the space of solutions of the classical equations as
suggested by \cite{wit}. By this way, starting from an arbitrary
Lagrangian, the phase space $Z$ will follow from the manifold of
field configurations. Our next step is to define the fundamental
objects on $Z$ for geometrical construction of the phase space.
\subsection{Fundamental objects on the phase space}
We assume that the gravitational field equations are derived from
the variation of a generic local gravity action that is a
functional in metric $g$, Riemann tensor $R$ and/or its contraction
and covariant derivatives\footnote{For the sake of simplicity, we
drop the indices on all tensorial quantities discussed in this
section.}
\begin{equation}
 S = \int  d^{D}x \, \sqrt{|g|} \, \mathcal{L}(g,R,\nabla R, R^{2},\cdots ).\label{S}
\end{equation}
Under first order variation, (\ref{S}) can be written as
\begin{equation}
  \delta S = \int  d^{D}x \, \sqrt{|g|} \,\Phi \,\delta g+ \int  d^{D}x
\, \partial\Lambda(g,\delta g,\nabla\delta g\cdots),\label{var}
\end{equation}
where $\Phi=0$ describes the field equations and $\partial\Lambda$
is a partial derivative of some boundary term with respect to the
spacetime coordinates.

Let $g$ be a solution of the field equations i.e. $\Phi(g)=0$. The
functions on $Z$, denoted by $g(x)$, take a spacetime point $x$ and
map it into a $D\times D$ real matrix $g(x)$. For the vectors,
consider an arbitrary and small variation in the metric $\tilde{g}=
g+\delta g$. When this is inserted into the field equations, it
yields $\tilde{\Phi}=\Phi+\delta\Phi$. Here $\delta\Phi$ are
obviously the linearized field equations. The vectors on $Z$ can be
defined as the variations $\delta g$ that solve $\delta \Phi=0$.
With vectors in hand, the corresponding differential one-forms are
easy to construct. A one-form, $\delta g(x)$, is the mapping from
the vector $\delta g$ to a $D\times D$ real matrix $\delta g(x)$,
which is the vector evaluated at a spacetime point $x$. We can
generalize this notion to construct more general $p$-forms as the
``wedge functions" of the one-forms $\delta g(x)$
\begin{equation}
 \Omega=\int dx_{1}\cdots dx_{p}\, \Theta(x_{1},\cdots ,x_{p})
\,\delta g(x_{1})\wedge\cdots\wedge\delta g(x_{p}),
\end{equation}
where $ \Theta(x_{1},\cdots ,x_{p})$ is a zero-form on $Z$ and
$\wedge$ is an anticommuting product. We can define an exterior
derivative operator $\delta$ that maps $p$-forms to ($p+1$)-forms
as follows
\begin{equation}
\delta \Omega=\int dx_{0}\, dx_{1}\cdots dx_{p}\,
\dfrac{\delta\Theta(x_{1},\cdots ,x_{p})}{\delta g(x_{0})} \,\delta
g(x_{0})\wedge\delta g(x_{1})\wedge\cdots\wedge\delta g(x_{p}),
\end{equation}
where $\dfrac{\delta\Theta(x_{1},\cdots ,x_{p})}{\delta g(x_{0})}$
is the functional derivative of $\Theta$ with respect to $g(x)$. One
can easily check that this operator obeys the modified Leibniz rule
and the celebrated Poincar\'{e} lemma. This construction and
notation was due to \cite{wit}, although one could also analyze the
same problem in a different approach  \cite{wald1,zuck}.
\subsection{The symplectic current and the symplectic two-form}
We are now ready to construct a symplectic two-form for the theories
described by (\ref{S}). First let us reconsider (\ref{var}) within
the context of the formalism we have reviewed in the previous
section. The variation of the action $\delta S$ can be viewed as a
one-form on $Z$ (note that $ \Lambda^{a}(x)$ contains $\delta g_{ab}$
and all of the other relevant quantities such as
$\delta{\Gamma^{a}_{bc}},\,\delta R_{ab},$ etc.). The key identity,
upon which the definition of covariantly conserved  symplectic
current is based, can be obtained from the exterior derivative of
(\ref{var}), which will vanish by the Poincar\'{e} Lemma
\begin{equation}
 \delta^{2}S= \int d^{D}x\,\sqrt{|g|}\,\delta \Phi_{ab} \wedge \delta g^{ab}
-\dfrac{1}{2}\int d^{D}x\,\sqrt{|g|}\, \Phi_{ab} \, \delta
g^{ab}\wedge \delta\lng +\int d^{D}x \,\partial_{a}\delta
\Lambda^{a}=0,\label{deltakare}
\end{equation}
where $\delta \lng= g^{ab}\delta g_{ab}=-g_{ab}\delta g^{ab}$. The
first two integrals vanish on-shell and the third one implies that
\begin{equation}
-\delta^{2}S=\int d^{D}x \,\sqrt{|g|}\,\nabla_{a}J^{a}=0,\label{symp}
\end{equation}
where $J^{a}\equiv -\delta\Lambda^{a}/\sqrt{|g|}$ is the
``symplectic current". We emphasize that this result holds on-shell
for any theory derived from (\ref{S}) and clarifies the definitions
of $J^{a}$ given in \cite{wit,bayramabi}.

Using (\ref{symp}), one can construct the following Poincar\'{e}
invariant two-form since the covariant divergence of the symplectic
current vanishes ($\nabla_{a}J^{a}=0$)
\begin{equation}
 \omega = \int_{\Sigma} d\Sigma_{a} \, \sqrt{|g|} \ J^{a } \,  \label{2form},
\end{equation}
where $\Sigma$ is a ($D-1$)-dimensional spacelike hypersurface.
Darbaoux's theorem guarantees that this is the sought-after
symplectic two-form of the theory if $\omega$ is also closed, which
can be shown by taking the exterior derivative of (\ref{2form})
\begin{equation}
 \delta \omega = \int_{\Sigma} d\Sigma_{a} \, (\delta\sqrt{|g|} \wedge J^{a } +\sqrt{|g|} \, \delta J^{a }). \label{closed}
\end{equation}
To evaluate the second term in (\ref{closed}), we now appeal to the
exterior derivative of (\ref{deltakare})
\begin{eqnarray}
\delta^{3}S&=&\int d^{D}x\,\sqrt{|g|}\,( \dfrac{1}{2}\, \delta\lng
\wedge \delta \Phi_{ab}\wedge\delta g^{ab} -
\dfrac{1}{2}\,\delta \Phi_{ab}\wedge  \delta g^{ab}\wedge \delta\lng \,)\\
&&+\int d^{D}x\,\sqrt{|g|}\,(\,\dfrac{1}{2}\, \delta\lng \wedge
\nabla_{a}J^{a}+\delta(\nabla_{a}J^{a})\,)=0,\nonumber
\end{eqnarray}
the first two terms cancel each other. Thus we obtain
\begin{equation}
\delta^{3}S=\dfrac{1}{2}\int d^{D}x\,\sqrt{|g|}\, \delta\lng \wedge
\nabla_{a}J^{a}+\int d^{D}x \,\sqrt{|g|}( \,\nabla_{a}\delta J^{a} +
\,\delta\Gamma^{b}\,_{ab}\wedge J^{a}\,)=0,
\end{equation}
from which an important relation follows
\begin{equation}
\int_{\Sigma} d\Sigma_{a}\,\sqrt{|g|}\,   \delta
J^{a}=-\dfrac{1}{2}\int_{\Sigma} d\Sigma_{a}\,\sqrt{|g|}\,
\,J^{a}\wedge\delta\lng.\label{virt}
\end{equation}
By virtue of (\ref{virt}) and bearing in mind that $J^{a}$ is an
anticommuting two-form, we see that (\ref{closed}) vanishes. It should
be noted that this result holds without the use of the field
equations, unlike the vanishing covariant divergence of $J^{a}$
that is valid only on-shell. This result was obtained for general
relativity and TMG by means of detailed calculations
\cite{wit,bayramabi}. Here we have given a completely general proof
applicable to the current $J^{a}$ derived from any local action
(\ref{S}).
\subsection{The gauge invariance}
Finally, one must show that the symplectic two-form is also gauge
invariant in the space of classical solutions $Z$ and in the
quotient space $\bar{Z}=Z/G$, where $G$ denotes the group of
diffeomorphisms ($x^{a}\rightarrow x^{a}+\xi^{a}$). The former is
trivial since all constituents of $\omega$ transform like tensors.
For the latter, we need to find out how $\omega$ transforms under
the following transformation
\begin{equation}
 \delta g_{a b} \rightarrow \delta g_{a b}+\nabla_{a}\xi_{b}+\nabla_{b}\xi_{a},\label{trans}
\end{equation}
where $\xi$ is asymptotic to a Killing vector field at the boundary
of the spacetime. Being a function of $\delta g_{ab}$,
 the transformation of one-forms will follow from (\ref{trans}) easily. Some of the basic quantities transform as
\begin{eqnarray}
 \delta\Gamma^{a}\,_{bc}& \rightarrow & \delta\Gamma^{a}\,_{bc}+ R_{ec}\,^{a}\,_{b}\,\xi^{e}+\nabla_{c}\nabla_{b}\xi^{a},\\
\delta R_{ab}& \rightarrow & \delta R_{ab}
+\xi^{c}\,\nabla_{c}R_{ab}+
R_{ad}\,\nabla_{b}\xi^{d}+R_{bd}\,\nabla_{a}\xi^{d}.
\end{eqnarray}
As a general rule for a tensor $T_{a\cdots}\,^{b\cdots}$, which is a
function of $\delta g_{ab}$ or its covariant derivatives, the
transformation reduces to
\begin{eqnarray}
\delta T_{a\cdots}\,^{b\cdots}& \rightarrow & \delta
T_{a\cdots}\,^{b\cdots} +\xi^{c}\nabla_{c}\,T_{a\cdots}\,^{b\cdots}+
T_{d\cdots}\,^{b\cdots}\,\nabla_{a}\xi^{d}+\cdots
-T_{a\cdots}\,^{d\cdots}\,\nabla^{b}\xi_{d}+\cdots\nonumber\\
&
=&\delta{T_{a\cdots}\,^{b\cdots}}+\mathcal{L}_{\xi}T_{a\cdots}\,^{b\cdots},
\end{eqnarray}
where $\mathcal{L}_{\xi}$ denotes the Lie derivative with respect to
the vector $\xi$. Note that this rule does not apply to Christoffel
symbols as they are not tensors. For $p$-forms, one should insert
the expressions above and keep those terms that are linear in
$\xi$. Then, the change in $\omega$ is given by
\begin{equation}
\Delta \omega = \int_{\Sigma} d\Sigma_{a} \, \sqrt{|g|} \ \Delta
J^{a }.\label{Omega}
\end{equation}
Now, if $\Delta J^{a }$ can be written as a divergence of an
antisymmetric two-form i.e. $\nabla_{a}\mathcal{F}^{ab}$ plus terms
that vanish on shell, $\omega$ is gauge invariant. The general proof
for an generic gravity theory derived from an action with local
symmetries is given in a corollary of \cite{wald1}. However, for our
purposes we will need the explicit form of $\mathcal{F}^{ab}$ to
define the conserved charges of e.g. the NMG theory.
\subsection{Application to $ {\mathcal L} \equiv \kappa^{-1} (R + 2 \Lambda_{0}) + \alpha R^{2} + \beta R_{ab}^{2} $}
We are now ready to apply this procedure to the following quadratic
action
\begin{equation}
 I = \int \, d^{D}x \, \sqrt{|g|} \, {\mathcal L} \equiv \int \, d^{D}x \, \sqrt{|g|} \, \Big( \frac{1}{\kappa} (R + 2 \Lambda_{0})
+ \alpha R^{2} + \beta R_{ab}^{2}  \Big).\label{eqm}
\end{equation}
The variation of (\ref{eqm}) reads
\begin{equation}
 \delta I = \int \, d^{D}x \, \sqrt{|g|} \,( \frac{1}{\kappa}{\mathcal G}_{ab}+\alpha  A_{ab}+ \beta  B_{ab}) \,\delta g^{ab}
+\int \, d^{D}x \Big(\frac{1}{\kappa}
\partial_{a}\Lambda^{a}_{\kappa}+\alpha\partial_{a}\Lambda^{a}_{\alpha}
+ \beta\partial_{a}\Lambda^{a}_{\beta}\Big),
\end{equation}
where
\begin{eqnarray}
 {\mathcal G}_{ab} & \equiv & R_{ab} - \frac{1}{2} g_{ab} R - \Lambda_{0} g_{ab}, \\
 A_{ab} & \equiv & 2 R R_{ab} - 2 \nabla_{a} \nabla_{b} R + g_{ab} (2 \square R - \frac{1}{2} R^2), \\
 B_{ab} & \equiv & 2 R_{acbd} R^{cd} - \nabla_{a} \nabla_{b} R +  \square R_{ab} + \frac{1}{2} g_{ab} (\square R - R_{cd} R^{cd}).
\end{eqnarray}
As discussed before, the boundary terms
\begin{eqnarray}
 \Lambda^{a}_{\kappa} & \equiv & \sqrt{|g|} \Big( g^{bc} \delta{\Gamma^{a}_{bc}}- g^{ab} \delta{\Gamma^{c}_{bc}}\Big), \\
\Lambda^{a}_{\alpha} & \equiv & \sqrt{|g|}\Big(2 R
g^{bc}\delta{\Gamma^{a}_{bc}}- 2 R g^{ab}\delta{\Gamma^{c}_{bc}}
+ 2 \nabla^{a}R \,\delta \lng + 2 \nabla_{b}R \,\delta g^{ab}\Big), \, \\
\Lambda^{a}_{\beta} & \equiv & \sqrt{|g|}\Big( 2 R^{cb}
\,\delta{\Gamma^{a}_{bc}}-  2 R^{ab}
\delta{\Gamma^{c}_{bc}}+\frac{1}{2} \nabla^{a}R \,\delta \lng +2
\nabla_{c}R^{a}\,_{b} \,\delta g^{bc}-\nabla^{a}R_{cb} \,\delta
g^{cb}  \Big),
\end{eqnarray}
yield a symplectic current given by
\begin{eqnarray}
 &J^{a}&= J^{a}_{\kappa}+J^{a}_{\alpha}+J^{a}_{\beta}\label{cur}, \quad \mbox{with}\\
 J^{a}_{\kappa}  =-\dfrac{\delta{\Lambda^{a}_{\kappa}}}{\sqrt{|g|}}&= &\delta{\Gamma^{a}_{bc}}\wedge(\delta g^{bc}+\frac{1}{2}g^{bc}\,\delta \lng)
-\delta{\Gamma^{c}_{bc}}\wedge(\delta g^{ab}+\frac{1}{2}g^{ab}\,\delta \lng), \\
J^{a}_{\alpha} =-\dfrac{ \delta{\Lambda^{a}_{\alpha}}}{\sqrt{|g|}}& =& \delta{\Gamma^{a}_{bc}}\wedge(2 R\, \delta g^{bc}+R g^{bc}\,\delta \lng+ 2 g^{bc}\,\delta R)\nonumber\\
&&-\delta{\Gamma^{c}_{bc}}\wedge(2 R\, \delta g^{ab}+R g^{ab}\,\delta \lng+ 2 g^{ab}\,\delta R) \\
&&-\delta \lng\wedge \Big(\nabla_{b}R\, \delta g^{ab}-2 \delta(\nabla^{a}R)\Big)+\delta g^{ab}\wedge\Big(2 \delta(\nabla_{b}R)-2\nabla_{b}R\,\delta \lng\Big) ,\nonumber\\
J^{a}_{\beta} =-\dfrac{ \delta{\Lambda^{a}_{\beta}}}{\sqrt{|g|}}&=&
\delta{\Gamma^{a}_{bc}}\wedge( R^{bc}\,\delta \lng+ 2 \delta R^{bc})
- \delta{\Gamma^{c}_{bc}}\wedge( R^{ab}\,\delta \lng+2 \delta R^{ab})\nonumber\\
&&+\delta \lng \wedge (\frac{1}{2} \delta(\nabla^{a}R)
- \nabla_{c}R^{a}\,_{b} \, \delta g^{bc} +\frac{1}{2} \nabla^{a}R_{cb}\,\delta g^{bc}) \\
&&+\delta g^{bc}\wedge\Big(2 \delta(\nabla_{c}R^{a}\,_{b})-
\delta(\nabla^{a}R_{cb})\Big).\nonumber
\end{eqnarray}
Here, the variation of several terms such as
$\delta(\nabla_{c}R^{a}\,_{b})$ are quite complicated and we save
the details to the Appendix. The covariant divergence of (\ref{cur})
vanishes on-shell as we discussed in the previous section.

There remains to investigate the gauge invariance of $\omega$. After
a cumbersome calculation, the change in the symplectic current can
be written as (the transformation properties of the relevant terms
are also given in the Appendix)
\begin{equation}
\Delta J^{a}=  \nabla_{c} {\mathcal F}^{ac} +2 \Phi_{bc} \,
\xi^{c}\wedge\delta g^{ab} + \Phi^{a}\,_{c}\,\xi^{c}\wedge\delta
\lng +\Phi_{bc}\, \xi^{a}\wedge \delta g^{bc}  , \label{change}
\end{equation}
where
\begin{equation}
 {\mathcal F}^{ac} = - {\mathcal F}^{ca} =  \frac{1}{\kappa} {\mathcal F}^{ac}_{\kappa} + \alpha {\mathcal F}^{ac}_{\alpha}
 + \beta {\mathcal F}^{ac}_{\beta}  \,, \label{deltaj}
\end{equation}
with
\begin{eqnarray}
{\mathcal F}^{ac}_{\kappa} & \equiv & 2  \xi^{[c}\wedge
\nabla_{b}\delta g^{a]b}-2 \xi_{b}\wedge \nabla^{[c}\delta
g^{a]b}-2\delta g^{b[c}\wedge\nabla_{b}\xi^{a]}
\nonumber\\& &-2\xi^{[a}\wedge\nabla^{c]}\delta \lng -\delta\lng \wedge \nabla^{[c}\xi^{a]},\\
{\mathcal F}^{ac}_{\alpha} & \equiv & 2 R {\mathcal F}^{ac}_{\kappa}
+4 \delta g^{b[c}\wedge \xi^{a]}\nabla_{b}R +4 \delta R\wedge
\nabla^{[a}\xi^{c]}
+ 8 \nabla^{[c}\delta R\wedge \xi^{a]},\\
{\mathcal F}^{ac}_{\beta}&\equiv & 2
R^{b[a}\delta\lng\wedge\nabla_{b}\xi^{c]}+4\,g^{d[a}\delta
R_{de}\wedge \nabla^{|e|}\xi^{c]}+2\,\delta\lng\wedge \xi^{b}
\nabla^{[c}R^{a]}\,_{b}
+4\,\delta g^{d[a}\wedge \nabla_{b}\xi^{c]} R_{d}\,^{b}\nonumber \\
& &-4 \,R_{e}\,^{[a} \nabla_{b}\xi^{c]}\wedge \delta g^{be} +4
R^{bd}\xi^{[a}\wedge \delta\Gamma^{c]}\,_{bd}
+4 R^{b[a}\xi^{c]}\wedge \delta\Gamma^{d}\,_{bd}-4\xi^{b}\wedge \delta g^{d[a}\nabla^{c]}R_{db}\nonumber \\
& &-4\xi_{e}\wedge \delta g^{b[c}\nabla_{b}R^{a]e} +4
g^{d[a}g^{c]e}\delta(\nabla_{e}R_{db})\wedge\xi^{b}-2\xi^{[a}\wedge\nabla^{c]}\delta
R
+2\delta g^{b[c}\wedge\xi^{a]}\nabla_{b}R  \\
& &-2 g^{bd} \xi^{[c}\wedge \nabla^{a]}\delta R_{bd}+4
g^{e[a}\xi^{c]}\wedge \nabla^{b}\delta R_{be}+4 g^{be}
R_{d}\,^{[c}\xi^{a]}\wedge \delta\Gamma^{d}\,_{be}
+4R^{d[a}\delta\Gamma^{c]}\,_{bd}\wedge \xi^{b}.\nonumber
\end{eqnarray}
The first term in (\ref{change}) vanishes when inserted in the
integral in (\ref{2form}) for sufficiently fast decaying metric
variations and the remaining terms vanish on-shell. In the next
section, we will discuss how conserved charges can be obtained from
(\ref{deltaj}) and will derive  an equality relating the ADT charge
definition  \cite{des1,des2,des3} and the charge expression obtained
from the symplectic two-form.
\section{\label{conserved}The conserved charges}
In a recent work \cite{bayramabi}, the conserved charges of the TMG
were obtained from the change in the symplectic current given in
(\ref{Omega}) under the group of diffeomorphisms. Here we use the
same idea to show that the charge expressions obtained in this
formalism and the ADT charge \cite{des1,des2,des3,biz} are
equivalent for theories derived from a local gravity action. We
first consider the transformation of (\ref{deltakare}) under
(\ref{trans})
\begin{eqnarray}
&&-2 \int d^{D}x\,\sqrt{|g|}\,\delta \Phi_{ab} \wedge
\nabla^{a}\xi^{b} +\int
d^{D}x\,\sqrt{|g|}\,\mathcal{L}_{\xi}\Phi_{ab} \wedge \delta g^{ab}
-\dfrac{1}{2}\int d^{D}x\,\sqrt{|g|}\, \Phi_{ab} \, \delta g^{ab}\wedge \nabla_{c}\xi^{c} \nonumber\\
& &+\int d^{D}x\,\sqrt{|g|}\, \Phi_{ab} \nabla^{a}\xi^{b} \wedge
\delta\lng -\int d^{D}x\, \sqrt{|g|}\,\nabla_{a}(\Delta
J^{a})=0\label{deltakaretrans},
\end{eqnarray}
where $\Delta J^{a}$ is the change in symplectic current. The first
term in (\ref{deltakaretrans}) can be cast as a divergence since
$\nabla_{a}{\delta \Phi^{ab}}=0$, which follows from the Bianchi
identity. Thus, we obtain
\begin{eqnarray}
 \int d^{D}x\,\sqrt{|g|}\,\nabla_{a}(2\,\delta \Phi^{ab} \wedge \xi_{b}+\Delta J^{a})
&=&\int d^{D}x\,\sqrt{|g|}\, \Phi_{ab} \nabla^{a}\xi^{b} \wedge \delta\lng \nonumber\\
&&-\int d^{D}x\,\sqrt{|g|}\,\delta g^{ab}\wedge(
\mathcal{L}_{\xi}\Phi_{ab}+\dfrac{1}{2}\,\Phi_{ab}
\,\nabla_{c}\xi^{c} ) \label{deltakaretrans2}.
\end{eqnarray}
We now further restrict our attention to the case where the metric
is linearized as $g_{ab}=\bar g_{ab}+h_{ab}$, and the deviation
$h_{ab}$ should vanish sufficiently slow as one approaches the
background $\bar g_{ab}$ at ``infinity"\footnote{Here this condition
guarantees nonzero results for the conserved charge.}. We also
assume that the background spacetime $\bar g_{ab}$ admits a globally
defined Killing vector $\bar \xi_{a}$. Indices are raised/lowered
and covariant derivatives are defined with respect to the background
metric $\bar g_{ab}$ as usual. The variation is identified as
$\delta g_{ab}\rightarrow h_{ab},\, \delta g^{ab}\rightarrow
-h^{ab}$. Therefore, the terms like $R_{ab},\, R$ are identified
with the  background ones $\bar R_{ab},\, \bar R$ and the terms like
$\delta(\nabla_{a}R_{bc})$ are taken as $(\nabla_{a}R_{bc})_{L}$,
where subscript $L$ indicates the linearized version of the
corresponding quantity. Finally, we put all of the $\xi$ terms into
the right hand side of the wedge products and then drop them. With
all of these identifications and by the help of field equations,
(\ref{deltakaretrans2}) yields
\begin{equation}
 \int d^{D}x\,\sqrt{|\bar g|}\,\bar\nabla_{a}((\Phi^{ab})_{L}  \bar\xi_{b}) = -\frac{1}{2}\int d^{D}x\,\sqrt{|\bar g|} \,\bar\nabla_{a}(\Delta \tilde{J}^{a}).\label{ADT}
\end{equation}
where $\Delta \tilde{J}^{a}$ is the vector obtained from the two-form
$\Delta J^{a}$ after identifications. The left hand side of
(\ref{ADT}) is the conserved current which is used to construct the
ADT \cite{des2} charge.\footnote{This relation accounts for the
factor of $-1/2$ used in \cite{bayramabi}.} From this we obtain the
charge expression as\footnote{$\Sigma$ is a $(D-1)$-dimensional
spacelike hypersurface with induced metric $\sigma$ and unit normal
vector $n^{a}$, $\partial\Sigma$ (boundary of $\Sigma$) is a
$(D-2)$-dimensional hypersurface with induced metric
$\sigma^{(\partial \Sigma)}$ and unit normal $s^{c}$. }
\begin{equation}
 Q_{ADT}(\bar\xi)=-\frac{1}{2} \int_{\Sigma} d^{D-1}x \, \sqrt{|\sigma|} \, n_{a} \,\bar{\nabla}_{c} { Q}^{ac} =
- \frac{1}{2} \int_{\partial \Sigma} d^{D-2}x \,
\sqrt{|\sigma^{(\partial \Sigma)}|} \, n_{a} \, s_{c} \, {Q}^{ac}
\,, \label{stokes}
\end{equation}
where
\begin{eqnarray}
  Q^{ac} &= &- Q^{ca} =  \frac{1}{\kappa}  Q^{ac}_{\kappa} + \alpha Q^{ac}_{\alpha}
 + \beta  Q^{ac}_{\beta}  \,,\\
Q^{ac}_{\kappa} & \equiv & 2   \bar\nabla_{b}h^{b[a}\bar\xi^{c]}-2
\bar\nabla^{[c}h^{a]b}\bar\xi_{b} +2 h^{b[c}
\bar\nabla_{b}\bar\xi^{a]}
+2(\bar\nabla^{[c}h) \bar\xi^{a]} -h \bar\nabla^{[c}\bar\xi^{a]},\label{kappa}\\
Q^{ac}_{\alpha} & \equiv & 2 \bar R Q^{ac}_{\kappa} - 4
\bar\nabla_{b}\bar R\, h^{b[c}\bar\xi^{a]} +4R_{L}
\bar\nabla^{[a}\bar\xi^{c]}
+ 8 (\bar\nabla^{[c}R_{L})\bar\xi^{a]},\label{alpha}\\
Q^{ac}_{\beta}&\equiv & 2 \bar
R^{b[a}h\bar\nabla_{b}\bar\xi^{c]}+4\,\bar
g^{d[a}(R_{de})_{L}\bar\nabla^{|e|}\bar\xi^{c]}
+2\,\bar\nabla^{[c}\bar R^{a]}\,_{b}  h \bar\xi^{b}
-4\,h^{d[a}\bar\nabla_{b}\bar\xi^{c]} \bar R_{d}\,^{b}\nonumber \\
& &+4 \,\bar R^{e[a} h_{be} \bar\nabla^{c]}\xi^{b} -4 \bar
R^{bd}(\Gamma^{[c}\,_{bd})_{L} \bar\xi^{a]}
-4 \bar R^{b[a} (\Gamma^{|d|}\,_{bd})_{L}\bar\xi^{c]} -4h^{d[a}\bar\xi^{|b|} \bar\nabla^{c]}\bar R_{db}\nonumber \\
& &-4 h^{b[c}\bar\xi_{e} \bar\nabla_{b}\bar R^{a]e} +4 \bar
g^{d[a}\bar g^{c]e}(\nabla_{e}R_{db})_{L}\bar\xi^{b}+2
(\bar\nabla^{[c}R_{L}) \bar\xi^{a]}
-2 h^{b[c} \bar\xi^{a]}\bar\nabla_{b}\bar R  \\
& &+2 \bar g^{bd}  \bar\nabla^{[a}(R_{bd})_{L}\,\bar\xi^{c]}-4 \bar
g^{e[a}\bar\nabla^{|b|}(R_{be})_{L}\bar\xi^{c]} -4 \bar g^{be} \bar
R_{d}\,^{[c} (\Gamma^{|d|}\,_{be})_{L} \bar\xi^{a]} +4\bar
R^{d[a}(\Gamma^{c]}\,_{bd})_{L} \bar\xi^{b}.\nonumber
\end{eqnarray}
The first two of the charge expressions (\ref{kappa}) and
(\ref{alpha}) are identical to their counterparts given in
\cite{biz},
 the equivalence of the third one can be shown after some computation. The next section is devoted to  the calculation of
the conserved charges of some solutions of NMG using this expression. 

\section{\label{soln}The conserved charges of some solutions of NMG}
Having found the charge expression (\ref{stokes}), let us consider
some black hole solutions of NMG for which we can use (\ref{stokes})
to compute the conserved charges. First we work out the examples
that are asymptotically AdS$_{3}$, e.g. the BTZ blackhole \cite{btz}
and the solutions given in \cite{clem3,ott}. Then we consider the
solutions with asymptotes that are not spaces of constant curvature,
namely the three-dimensional Lifshitz black hole \cite{giri1} and the warped
AdS$_{3}$ black hole given in \cite{clem}. Both examples have been
studied in \cite{biz2,clem,toni} with which we compare the results.
\subsection{The BTZ black hole}
The first example  is the celebrated BTZ black hole \cite{btz},
which can be cast in the form
\begin{equation}
 ds^2 \ = \Big(\frac{-2\rho}{l^{2}}+\frac{M}{2}\Big) dt^{2}+\Big(\frac{4 \rho^{2}}{l^{2}}
-\frac{(M^{2} l^{2} -J^{2})}{4}\Big)^{-1} d\rho^2-J dt d\phi
+\Big(2\rho+\frac{Ml^{2}}{2}\Big) d\phi^{2}\label{BTZ},
\end{equation}
and this is a solution of NMG when
\begin{equation}
 \kappa=16\pi G,\,\,\beta=-\frac{1}{\kappa m^{2}},\,\,\Lambda_{0}=\frac{1+4l^{2}m^{2}}{4 l^{4}m^{2}},\,\,\alpha=-\frac{3}{8}\beta.
\end{equation}
Here $m^{2}$ is a ``relative'' mass parameter of the NMG
\cite{hohm}. The background spacetime is taken to be AdS$_{3}$ that
is obtained by setting $M\rightarrow0$, $J\rightarrow0$ in
(\ref{BTZ})
\begin{equation}
ds^{2}=-\frac{2\rho}{l^{2}}dt^{2}+\frac{l^{2}}{4\rho^{2}}d\rho^{2}+2\rho
d\phi^{2}.\label{back}
\end{equation}
This form of AdS$_{3}$ clearly possesses two globally defined
Killing vectors $\bar\xi^{a}=(-1,0,0)$ and
$\bar\vartheta^{a}=(0,0,1)$ that are used in the computation of the
energy and angular momentum, respectively. The timelike and spacelike
normals that follow from the normalization condition i.e.
$n^{a}n_{a}=-1$, $s^{a}s_{a}=+1$ are
 \[n^{a}=-\frac{\ell}{\sqrt{2 \rho}}\delta^{a}_{t},\,\, s^{a}=\frac{2 \rho}{\ell}\delta^{a}_{\rho}. \]
Finally the measure of (\ref{stokes}) is simply
$\sqrt{|\sigma^{(\partial \Sigma)}|}=\sqrt{2\rho}$. The conserved
charges are obtained using (\ref{stokes})
\begin{equation}
E_{BTZ}=\Big(1-\frac{1}{2
l^{2}m^{2}}\Big)\frac{M}{16G},\,\quad\,J_{BTZ}=\Big(1-\frac{1}{2
l^{2}m^{2}}\Big)\frac{J}{16G}.
\end{equation}
These ``renormalized mass and angular momentum'' coincide with the
ones given in \cite{clem} that employed ADT charge definition for
computation and \cite{toni} in which the boundary stress tensor method
was used.
\subsection{The ``logarithmic'' black hole in \cite{clem3}}
As a second example consider the black hole solution given in
\cite{clem3}
\begin{equation}
 ds^2
= -\frac{4\rho^2}{\ell^2f(\rho)}d{t}^2 + f(\rho)\bigg[ d\phi -
\frac{q\ell\ln|\rho/\rho_0|}{f(\rho)}d{t}\bigg]^2 +
\frac{\ell^2d\rho^2}{4\rho^2} ,
\end{equation}
where
\[f(\rho) = 2\rho+q\ell^2\ln|\rho/\rho_0|.\]
This is a solution to the NMG with
\begin{equation}
 \kappa=8\pi G,\,\,\beta=-\frac{2\ell^{2}}{\kappa},\,\,\Lambda_{0}=\frac{3}{2\ell^{2}}.
\end{equation}
The background spacetime is taken to be AdS$_{3}$ in the form
(\ref{back}), therefore we can employ the same Killing vectors,
normals and induced metric as in the BTZ case. Following the same
lines for the calculation of charges, we find
\begin{eqnarray}
 E&=&\lim_{\rho \to \infty} \int_{0}^{2 \pi } \sqrt{2\rho}\,  n_{t} \, s_{\rho} \, {Q}^{t\rho}(\bar\xi) d\phi=\frac{2q}{G},\\
J&=&\lim_{\rho \to \infty} \int_{0}^{2 \pi } \sqrt{2\rho}\,  n_{t}
\, s_{\rho} \, {Q}^{t\rho}(\bar\vartheta) d\phi =\frac{2\ell q}{G}.
\end{eqnarray}
This result is identical to the one given in \cite{clem3} that was
again computed through ADT.
\subsection{The rotating black hole in \cite{ott}}
The next example that is of interest is the stationary solution
given in \cite{ott}
\begin{equation}
 ds^{2}=\Big(-N(r) F(r) +r^{2} K(r)^{2}\Big) \,dt^{2}+\dfrac{dr^{2}}{F(r)} +2r^{2}K(r) \,dt\,d\phi +r^{2}\,d\phi^{2}\label{ottmet},
\end{equation}
where
\begin{eqnarray}
 N(r)&=&\Big[1+\frac{q \ell^{2}}{4 H(r)}(1-\sqrt{\Xi})\Big]^{2},\\
F(r)&=&\frac{H(r)^{2}}{r^{2}}\Big[\frac{H(r)^{2}}{\ell^{2}}+\frac{q}{2}(1+\sqrt{\Xi})H(r)+\frac{q^{2}\ell^{2}}{16}(1-\sqrt{\Xi})^{2}-4GM\sqrt{\Xi}\Big],\\
K(r)&=&-\frac{p}{2r^{2}}(4GM-qH(r)),\\
H(r)&=& \Big[r^{2}-2GM \ell^{2}(1-\sqrt{\Xi})-\frac{q^{2}\ell^{4}}{16}(1-\sqrt{\Xi})^{2}\Big]^{1/2},\\
\Xi&\equiv &1-p^{2}/\ell^{2},
\end{eqnarray}
with
\[\Lambda_{0}=\frac{1}{2\ell^{2}}, \quad \beta=\frac{2\ell^{2}}{\kappa},\,\,\alpha=-\frac{3}{8}\beta
,\,\, \quad \kappa=16\pi G.\] in our conventions. The rotation
parameter $p$ is restricted between $-\ell \leq p\leq \ell$ and the
parameter $q$ is the additional ``gravitational hair'' for which the
$b=0$ case is the BTZ blackhole. The background spacetime relevant
for our purposes can be found by setting
 $q\rightarrow0$, $M\rightarrow0$ in (\ref{ottmet}) that is simply AdS$_{3}$ spacetime
\begin{equation}
ds^{2}=-\frac{r^{2}}{l^{2}}dt^{2}+\frac{l^{2}}{r^{2}}dr^{2}+r^{2}d\phi^{2}.
\end{equation}
The timelike and spacelike normals are
\[n^{a}=-\frac{\ell}{r}\delta^{a}_{t},\,\, s^{a}=\frac{r}{\ell}\delta^{a}_{r},\,\,\sigma^{(\partial \Sigma)} = r^{2} .\]
With those choices we compute the energy and angular momentum to be
\begin{eqnarray}
 E &=& \lim_{r \to \infty} \int_{0}^{2 \pi } r\,  n_{t} \, s_{r} \, {Q}^{tr}(\bar\xi) d\phi = M,\\
J &=&\lim_{r \to \infty} \int_{0}^{2 \pi } r\, n_{t} \, s_{r} \,
{Q}^{tr}(\bar\vartheta) d\phi = M p.
\end{eqnarray}
As discussed in \cite{ott}, the parameter $b$ does not appear in the
conserved charges, which is the reason it was called 
``gravitational hair'' in the first place.
\subsection{Three-dimensional Lifshitz black hole}
The first example with a nonconstant curvature background is the three-dimensional 
Lifshitz black hole \cite{giri1} given as
\begin{equation}
 ds^2 = - \frac{r^6}{\ell^6} \,\Big(1 - \frac{M \ell^2}{r^2}\Big)\, dt^2 + \frac{\ell^{2}}{r^{2}}\Big(1 - \frac{M \ell^2}{r^2}\Big)^{-1}\, dr^2+ \frac{r^2}{\ell^2} dx^2,
 \label{lif3d}
\end{equation}
which solves NMG with
\[ \Lambda_{0} = \frac{13}{2 \ell^2}, \quad \beta = \frac{2 \ell^2}{\kappa}, \quad \alpha = - \frac{3 \ell^2}{4 \kappa},\quad \kappa=16 \pi G.\]
The background metric can be obtained by taking $M\rightarrow 0$
\[ ds^2 = - \frac{r^6}{\ell^6} dt^2 + \frac{\ell^2}{r^2} dr^2 + \frac{r^2}{\ell^2} dx^2. \]
The timelike, spacelike normals and one-dimensional induced metric can
easily be found as
\[n^{a}=-\frac{\ell^{3}}{r^{3}}\delta^{a}_{t},\,\, s^{a}=\frac{r}{\ell}\delta^{a}_{r},\,\,\sigma^{(\partial \Sigma)}=\dfrac{r^{2}}{\ell^{2}}.\]
For the energy, the timelike Killing vector
$\bar\xi^{a}=-\delta^{a}_{t}$ can be employed. With these,
(\ref{stokes}) yields
\begin{equation}
 E = \lim_{r \to \infty} \int_{0}^{2 \pi \ell} \frac{r}{\ell}\, n_{t} \, s_{r} \, {Q}^{tr}(\bar\xi) dx =\frac{7 M^2}{8G}. \label{enlif}
\end{equation}
That is the same energy given in \cite{biz} that was calculated
through the ADT procedure for arbitrary backgrounds, yet the result
differs from the expression in \cite{toni}.
\subsection{The Warped AdS$_{3}$ black hole}
The final example is the warped AdS$_{3}$ black hole \cite{clem}
that reads
\begin{equation}
ds^{2}= -\mu^{2}
\dfrac{r^{2}-r_{0}^{2}}{F(r)}\,dt^{2}+F(r)\Big[d\phi -
\dfrac{r+(1-\mu^{2})\omega}{F(r)} \,dt\Big]^{2}
+\frac{1}{\mu^{2}\zeta^{2}}\,\dfrac{dr^{2}}{r^{2}-r_{0}^{2}},
\label{warped}
\end{equation}
where
\[ F(r)=r^{2}+2\omega r+\omega^{2}(1-\mu^{2})+\dfrac{\mu^{2}r^{2}_{0}}{1-\mu^{2}}. \]
This is a solution of the NMG theory with
\[ \kappa = 8 \pi G ,\, \quad \beta = - \frac{1}{m^2 \kappa} \,, \quad
 \alpha  = \frac{3}{8 m^2 \kappa}, \]
\[ \mu^2 = \frac{9 m^2 + 21 \Lambda_{0} - 2 m \sqrt{3 (5 m^2 - 7 \Lambda_{0})}}{4 (m^2 + \Lambda_{0})}
\quad \mbox{and} \quad \zeta^2 = \frac{8 m^2}{21 - 4 \mu^2} \] with
$m^{2}$ as the NMG parameter. In order to have a causally regular
black hole, $\mu^{2}$ and $\Lambda_{0}$ must be \cite{clem}
\[ 0 < \mu^2 < 1 \quad \mbox{and} \quad \frac{35 m^2}{289} \geq \Lambda_{0} \geq - \frac{m^2}{21} \,. \]
The background spacetime of this black hole can be defined by taking
$\omega \rightarrow 0,\,\, r_{0}\rightarrow 0$ in (\ref{warped})
\begin{equation}
 ds^{2}=(1-\mu^{2})\,dt^{2}+\dfrac{1}{r^{2}\zeta^{2}\mu^{2}}\,dr^{2}-2r\,d\phi\,dt+r^{2}d\phi^{2}.\label{warpedback}
\end{equation}
The timelike, spacelike normals and the measure is apparent
considering the standard ADM  form of the metric (\ref{warpedback})
\[n_{a}= -\mu \,\delta^{t}_{a},\quad s_{a}=\dfrac{1}{\mu r \zeta}\,\delta_{a}^{r},\,\,\sqrt{|\sigma^{(\partial \Sigma)}|} = r.\]
To find the energy, one again has to choose the timelike Killing vector
as $\bar\xi^{a}=-\delta^{a}_{t}$  and for the angular momentum one
has to use $\bar\vartheta^{a}=\delta^{a}_{\phi}$. Then
(\ref{stokes}) yields
\begin{eqnarray}
 E &= &\lim_{r \to \infty} \int_{0}^{2 \pi }  r\, n_{t} \, s_{r} \, {Q}^{tr}(\bar\xi) d\phi = \dfrac{4 \mu^{2}(1-\mu^{2})\omega \zeta}{G(21-4\mu^{2})},\\
J &=&\lim_{r \to \infty} \int_{0}^{2 \pi } r\, n_{t} \, s_{r} \,
{Q}^{tr}(\bar\vartheta) d\phi
=-\dfrac{\zeta}{8G(21-4\mu^{2})}\Big[\dfrac{16
r_{0}^{2}\mu^{2}}{(1-\mu^{2})}+\dfrac{(1-\mu^{2})}{\mu^{2}}(21-29\mu^{2}+24\mu^{4})\omega^{2}\Big].
\nonumber
\end{eqnarray}
The values for the energy and angular momentum agrees with the ones
given in \cite{biz2}, however angular momentum is in conflict with
the one in \cite{clem}. The discrepancy of these results, and the
validity of the charge expression are discussed more explicitly in
\cite{biz2}.

\section{Conclusions}
In this paper, starting from a local gravity action described by
(\ref{S}), we showed that a covariantly conserved symplectic current
can always be obtained from the boundary terms that appear in the
first variation of the action. Moreover, we have shown that the
two-form obtained from the integration of the symplectic current over
a spacelike hypersurface is closed for any theory. The investigation
of the gauge invariance of this two-form is the final task that needs
to be performed in order to show that it is the symplectic two-form of
the theory, which provides the most important result of this paper.
Under diffeomorphisms, the symplectic two-form of a generic gravity
theory  yields a conserved Killing charge expression that is
equivalent to the extended ADT formalism for arbitrary backgrounds
with at least one global Killing  isometry \cite{biz}.

As a consistency check, we obtained the charge expression for NMG
and calculated the energy and angular momentum of several black
holes. The charges of black holes with AdS$_{3}$ backgrounds are in
agreement with the previous works \cite{clem,toni,ott}. In the case
of nonconstant curvature backgrounds, namely, Lifshitz and warped
AdS$_{3}$ spacetimes, the results agree with the ones computed
through the ADT procedure for arbitrary backgrounds \cite{biz,biz2},
which was expected since the charge expressions were shown to be
covariantly equivalent. On the other hand, it was shown in
\cite{biz2} that in the case of Lifshitz and warped AdS$_{3}$ black
holes, the charge expressions obtained were in conflict with the
results found by other means \cite{clem,toni}. This discrepancy
calls for further study regarding the validity of this charge
expression for generic backgrounds. It would also be interesting to
perform a covariant, geometric quantization of the theories
described by (\ref{eqm}).

\appendix*

\section{\label{appa}}
Here we first present the identities that are used to compute the
variation of a covariant derivative of a tensor. Then, we list the
transformation of terms that we have used during the calculation of
$\Delta\omega$.

From the well known equality of Christoffel symbols
\begin{equation}
 \delta\Gamma^{a}\,_{bc} = \frac{1}{2} g^{ad}
 \left( \nabla_{b} \delta g_{cd} + \nabla_{c} \delta g_{bd} - \nabla_{d} \delta g_{bc} \right).
\end{equation}
variation of the Riemann tensor can be calculated simply
\begin{equation}
\delta R^{a}\,_{bcd} =  \nabla_{c} \delta\Gamma^{a}\,_{bd} -
\nabla_{d} \delta\Gamma^{a}\,_{bc},
\end{equation}
and the contraction of indices leads to
\begin{equation}
\delta R_{ab}=\nabla_{c} \delta\Gamma^{c}\,_{ab} -  \nabla_{b}
\delta\Gamma^{c}\,_{ac}.
\end{equation}
After a straightforward calculation the variation of the covariant
derivative of a tensor $T_{b\cdots}\,^{c\cdots}$ can be written as
\begin{equation}
\delta(\nabla_{a}T_{b\cdots}\,^{c\cdots})=\nabla_{a}\delta
T_{b\cdots}\,^{c\cdots}-  \delta\Gamma^{i}\,_{ab}\,
T_{i\cdots}\,^{c\cdots} -\cdots +\delta\Gamma^{c}\,_{ai}\,
T_{b\cdots}\,^{i\cdots}+\cdots,
\end{equation}
which is reminiscent of the usual covariant derivative formula where
Christoffel symbols are replaced with $\delta\Gamma^{a}\,_{bc}$.
Application of this formula together with the identities above leads
to the following useful relations
\begin{eqnarray}
\delta(\nabla_{a} R_{cd})& =& \nabla_{a} \delta R_{cd} -R_{ed}\, \delta\Gamma^{e}\,_{ac}- R_{ec}\,\delta\Gamma^{e}\,_{ad},\nonumber\\
\delta(\nabla_{b} \nabla_{a} R_{cd}) & = & \nabla_{b}
\delta(\nabla_{a} R_{cd})
 -   \nabla_{e} R_{cd}\, \delta\Gamma^{e}\,_{ba}
 - \nabla_{a} R_{ed} \, \delta\Gamma^{e}\,_{bc} - \nabla_{a} R_{ce}\, \delta\Gamma^{e}\,_{bd}, \nonumber \\
 \delta(\square R_{cd}) & = & g^{ab}\, \delta(\nabla_{b} \nabla_{a} R_{cd})
 +\nabla_{b} \nabla_{a} R_{cd}\, \delta g^{ab} , \nonumber \\
 \delta(\nabla_{b} \nabla_{a} R) & = & g^{cd} \delta(\nabla_{b} \nabla_{a} R_{cd})
 +\nabla_{b} \nabla_{a} R_{cd} \,\delta{g}^{cd} ,  \nonumber \\
 \delta(\square R) & = & g^{ab}\, \delta(\nabla_{b} \nabla_{a} R)
 + \nabla_{b} \nabla_{a} R\,\delta g^{ab}.
\end{eqnarray}
Throughout the calculation of $\Delta\omega$, the transformation of
the following terms under $\delta g_{a b} \rightarrow \delta g_{a
b}+\nabla_{a}\xi_{b}+\nabla_{b}\xi_{a}$ has also been used:
\begin{eqnarray}
 \delta\Gamma^{a}\,_{bc}& \rightarrow & \delta\Gamma^{a}\,_{bc}+ R_{ec}\,^{a}\,_{b}\,\xi^{e}+\nabla_{c}\nabla_{b}\xi^{a},\nonumber\\
\delta\lng& \rightarrow & \delta\lng +2\nabla_{a}\xi^{a},\nonumber\\
\delta R & \rightarrow & \delta R + \xi^{a}\nabla_{a}R,\nonumber\\
\delta(\nabla_{a}R) & \rightarrow & \delta(\nabla_{a}R)+ \nabla_{a}\nabla_{b}R\,\xi^{b}+\nabla_{b}R\, \nabla^{b}\xi_{a},\nonumber\\
\delta R_{ab}& \rightarrow & \delta R_{ab}
+\nabla_{c}R_{ab}\,\xi^{c}+
R_{ad}\,\nabla_{b}\xi^{d}+R_{bd}\,\nabla_{a}\xi^{d}.
\end{eqnarray}
As stated previously, the change in the variation of a tensor is
given by the Lie derivative of that tensor along $\xi$.
\begin{acknowledgments}
 We are grateful to {\"O}zg{\"u}r Sar{\i}o\u{g}lu for his guidance and critical reading of the manuscript. We also
 thank Bayram Tekin for his useful comments.
This work is partially supported by the Scientific and Technological
Research Council of Turkey (T{\"U}B\.{I}TAK). GA is supported by
T{\"U}B\.{I}TAK MSc Scholarship.
\end{acknowledgments}

\end{document}